# Atomic structural characteristics and dynamical properties in monatomic metallic liquids via molecular dynamics simulations

X. Qin, J.Q. Wu, and M. Z. Li[*]

*Department of Physics and Beijing Key Laboratory of Opto-Electronic Functional Materials & Micro-Nano Devices, Renmin University of China, Beijing 100872, China*

## Abstract

Molecular dynamics simulations were performed for five monatomic metallic liquids and the atomic structural characteristics and dynamical properties were systematically investigated and compared for understanding the underlying structural basis of liquid properties, such as glass-forming ability. All simulated monatomic liquids exhibit similar structural characteristics and temperature evolution. However, the degree varies significantly in different liquids. It is found that the atomic structures in liquid Cu, Ta, and Fe are quite similar. However, liquid Ta exhibits more ordered and more densely packed structure features, with more populated icosahedral-like atomic ordering and five-fold HA indexes, less Voronoi entropy, and more regular tetrahedral configurations. Moreover, the increase of crystal-like clusters in liquid Ta slows down with decreasing temperature, and the icosahedral-like clusters increase more quickly, exceeding crystal-like ones, in contrast to other liquids. On the contrary, the atomic structures of liquid Al and Zr are more similar, much more disordered and more loosely packed. Furthermore, liquid Ta exhibits more slow and heterogeneous dynamics, which could be facilitated by the particular atomic structures of liquid Ta. Both the atomic structure and dynamics features in liquid Ta favor its glass formation. Our findings provide comprehensive structural and dynamical information for better understanding GFA and crystallization in these typical monatomic liquids.

*maozhili@ruc.edu.cn

# I. INTRODUCTION

Glass is formed when a liquid is cooled rapidly to prevent crystallization. Since the successful synthesis of Au-Si metallic glass (MG) in 1960 [1], the search for new types of MGs and the mechanism of their formation has not stopped [2-4]. However, glass forming ability (GFA) of metallic alloys varies greatly with different components [5]. Doping atoms in alloys can be effective in improving GFA. For example, by doping Al, the supercooled liquids of Cu-Zr based alloys are more stable, with GFA significantly improved [6]. Doping Ta into bulk MGs can also effectively improve GFA and mechanical properties [7]. However, it has been generally believed that monatomic metals cannot form MGs in experiments.

Only recently have monatomic MGs been successfully obtained experimentally. The liquid Ta, V, W, and Mo of body-centered-cubic (BCC) crystal phase were rapidly cooled to form monatomic MG by using nanobridge cooling devices that a very high cooling rate of $10^{14}$ K/s can be achieved in this way [8]. However, for metals with face-centered-cubic (FCC) structure, such as Au, Ag, Cu, and Al, it is still impossible to successfully produce monatomic MG in experiments.

A lot of studies have been carried out, trying to understand GFA of monatomic liquid from structural characteristics [8-19]. In 1952, Frank proposed that the main structure of supercooled liquid of pure metal may be icosahedron which is incompatible with the long-range periodicity of crystal [14]. In addition, diffraction experiment has detected the existence of icosahedral short-range ordering in liquid Fe and Zr [15]. by computer simulations, other local structures were also revealed in monatomic liquids. By *ab initio* molecular dynamics (MD) simulations for liquid Zr, the competition of different local structures and the increase of BCC-like atomic structures in liquid Zr were observed [16]. This suggests that the local structural ordering is the result of the competition between BCC order and polyhedral order, as well as the fact that the liquid structure is much more complex than icosahedral order [16]. However, K. Nishio *et al*. performed *ab initio* MD simulations for various monatomic liquids and found that entropy-driven docosahedral local structures are dominant and independent of their crystal structures [17]. This suggests that the docosahedral short-range order could be the underlying structural basis of better GFA of BCC metals than FCC ones [17]. However, apart from docsahedron, various atomic clusters are also commonly populated in different monatomic metallic liquids, which might also play some roles in GFA of monatomic metallic liquids [18]. In order to investigate how different structures affect GFA of monatomic liquids, crystallization processes of liquid Cu and Zr were investigated [19]. It is found that in supercooled liquid states, crystalline structure-like fraction in liquid Cu is much higher than that of Zr, which reduces the interface energy between liquid and crystallographic regions and results in low GFA. Therefore, a lot of effort has been devoted to the important issue, however, the microscopic mechanism of glass formation in the monatomic system is not clear.

This paper attempts to explore and understand the GFA, microstructure and dynamics of five monatomic liquids by systematically studying and comparing their structural and dynamic evolutionary features. The results show that from the comparison of different elements, the structures of Cu and Fe are similar, the structures of Al and Zr are similar. Combined with icosahedron and HA index, we found a possible reason why the GFA of Cu is stronger than that of Al and that of Ta is stronger than that of Zr. In dynamics, it is found that during the cooling process, the liquid gradually breaks the Stokes-Einstein relation and the dynamic heterogeneity is enhanced and the dynamic behaviors of Cu and Ta are very similar.

## II. Model and Method

We performed classical MD simulations within the LAMMPS package to prepare and analyze monatomic metallic liquid Cu, Ta, Fe, Al, and Zr, respectively [20]. The interatomic interaction for each system was described by the realistic embedded-atom method potential [8, 21-23]. In MD simulations, each sample contains 40000 atoms in a cubic box with periodic boundary conditions applied in three dimensions. The sample was first fully equilibrated at high temperature above melting point, then fast quenched into supercooled liquid states with cooling rate of $10^{14}$ K/s. The quenching process was carried out in NPT ensemble with pressure adjusted to be zero [24,25]. At each temperature point of interest, isothermal annealing was performed for each sample in NVT ensemble for 1ns. Meanwhile, atomic configurations of each sample were collected for the analysis of structure and dynamics. The lowest temperature was taken at which crystallization was not observed in the isothermally annealed samples. In the above simulations, the MD time step was set as 2 fs. To make easy comparison among these monatomic liquids, temperature T was scaled by melting temperature $T_m$ in the following calculation. For Cu, Ta, Fe, Al, and Zr, the melting points are 1356K, 3266K, 933K, 2125K, and 1811K, respectively [26].

## III. Results and Discussions

### A. Radial distribution function analysis

First, we analyzed the radial distribution functions (RDF) in these monatomic metallic liquids (see Supplementary Materials S1). Typical behavior and temperature evolution of RDFs in liquids can be observed. With decreasing temperature, the first peak position shifts towards larger distance in all simulated liquids, as illustrated in Figure 1(a). Previous studies demonstrate that this mainly results from the increase of the coordination number (CN) as temperature decreases [27]. We also evaluated CN as a function of temperature by integrating RDF up to the first local minimum. As shown in Fig. 1(b), the CN increases with decreasing temperature in all simulated liquids. This implies that local atomic structures change with temperature in these liquids. It can be seen that CN in liquid Ta is the largest among these metallic liquids, whereas it is quite similar in liquid Cu, Zr, and Fe. This implies that the local atomic structures in liquid Ta could be quite different from those in other monatomic liquids.

We also compared the ratio of the first peak height to the first local minimum of RDFs in theses monatomic liquids as functions of temperature. The increase of the ratio means that short-range atomic structures become more ordered. As shown in Fig. 1(c), the ratio significantly increases with decreasing temperature in each liquid. However, the increase is the most in liquid Ta. Moreover, the ratio in liquid Cu and Fe shows similar temperature-dependent behavior. This indicates that the temperature evolution of local atomic ordering is more significant in liquid Ta.

Furthermore, the second peak also increases with decreasing temperature in these liquids. For liquid Cu, a shoulder peak is gradually developed. For other liquids, however, this feature is not clearly seen. This feature often refers to the development of medium-range order (MRO) [28]. In this sense, a significant MRO is developed in liquid Cu with decreasing temperature. As mentioned above, GFA of liquid Cu is not good among these simulated liquids. Therefore, the development of MRO in liquid Cu could have little relation to GFA.

**B. Voronoi tessellation analysis**

To characterize the local atomic structures in these simulated monatomic metallic liquids, Voronoi tessellation was employed, in which local atomic structures can be identified by Voronoi index $<n_3, n_4, n_5, n_6>$, where $n_i$ ($i$=3,4,5,6) denotes the number of $i$-edged faces in a Voronoi cluster [29]. For example, $<0,0,12,0>$ represents the icosahedral cluster. Figure 2 shows the temperature evolution of the fraction of some Voronoi clusters in each liquid. Here top 10 populated clusters in the lowest-temperature liquids were selected in each monatomic liquid. It can be seen that the selected cluster types are the same in liquid Cu and Fe. In liquid Al and Zr, the most populated clusters are the same, too. However, the fraction of the same clusters is different. In this sense, the atomic structures of liquid Cu and Fe are similar, and liquid Al and Zr share more similar atomic structure characteristics. In liquid Ta, the cluster types are slightly different from other monatomic liquids. In all these liquids, $<0,2,8,4>$, $<0,1,10,2>$, and $<0,3,6,4>$ clusters are most populated, which may be a general structural characteristic in monatomic liquids. However, the fraction is larger in liquid Ta. It is also found that fraction of the selected clusters in liquid Al is significantly lower than other liquids. In addition, no icosahedral cluster of <0,0,12,0> is observed in liquid Al, which may be associated with poor GFA of metal Al. The situation of Zr is similar, and the cluster fraction is not high, either.

To better illustrate evolution of structural ordering with decreasing temperature in the simulated monatomic liquids, the population of crystal-like and icosahedral-like clusters was compared. According to previous studies, <0,3,6,4>, <0,3,6,5>, <0,4,4,6>, <0,4,4,7>, <0,6,0,8>, and <0,5,2,8> are often considered as crystal-like clusters, and <0,0,12,0>, <0,1,10,2>, <1,0,9,3>, and <0,2,8,2> are classified as icosahedral like [16,30]. Figure 3 shows temperature evolution of fraction of crystal-like and icosahedral-like clusters in the simulated monatomic liquids, respectively. It can be seen that both crystal-like and icosahedral-like clusters increases with decreasing temperature in these monatomic liquids. Above melting temperature, the population of icosahedral-like clusters is lower than the crystal-like ones in all simulated liquids. For liquid Cu and Ta, however, the population of icosahedral-like clusters increases quickly below melting temperature, getting higher than the crystal-like clusters, especially in liquid Ta. Moreover, the increase of crystal-like clusters in liquid Zr gradually slows down with the decrease of temperature, in contrast to other monatomic liquids. These unique structure characteristics in liquid Ta favor the glass formation of Ta. Previous studies show that the population of crystal-like structures influences the interfacial energy of the crystallization in a liquid system [19]. This may explain why liquid Al and Zr is easily crystallized. For liquid Cu, however, it might not be the case.

To further characterize the evolution of structural ordering, Voronoi entropy was analyzed, which is defined as

$$S = -\sum_{i=1}^{N} p_i \ln p_i$$

where $N$ and $p_i$ represent the number of cluster type and the fraction of the Voronoi cluster of type $i$ [31,32]. Figure 4(a) shows the evolution of Voronoi entropy with decreasing temperature in these liquids. It can be seen that Voronoi entropy decreases with the decrease of temperature in all simulated liquids, indicating that the atomic structures of liquids become more ordered in cooling process. While liquid Zr shows larger entropy values, in liquid Ta the values are smaller. For liquid Cu, Al and Fe, entropy values locate in between, and they are almost the same in liquid Cu and Fe. This indicates that liquid Ta is more ordered than other monatomic liquids.

It can be seen that Voronoi entropy is determined by both the number of cluster type and cluster probability distribution. To get more insight into the temperature evolution of Voronoi entropy, we also analyzed the evolution of the number of cluster type and cluster probability distribution. Here the variance of the cluster fraction, $\sigma^2$ was adopted to evaluate cluster probability distribution in a liquid, which describes the difference in the relative contribution of cluster types and is defined as [33],

$$\sigma^2 = \frac{\sum_i (p_i - \mu)^2}{N}$$

where $p_i$ is the fractional contribution of a particular type of cluster to the structure, $\mu$ the mean of the cluster fraction, and $N$ the number of cluster types, respectively.

Figure 4(b) and 4(c) show the temperature evolution of the number of Voronoi cluster types and cluster variance, respectively. The more even the cluster probability distribution is, the smaller the variance is. It can be seen that the number of cluster type decreases with decreasing temperature. However, cluster variance increases as temperature decreases, implying that cluster fraction becomes much more different in cooling process. As shown in Figure 4(b-c), liquid Ta contains fewer cluster types, but exhibits larger variance values, compared to other monatomic liquids. On the contrary, liquid Zr possesses more types of clusters and smaller variance values. These results suggest that the structures of liquid Ta could be more ordered and more heterogenous.

Note that Voronoi entropy only reflects order degree of the system, however, cannot judge which order it is, such as amorphous order or crystal order. Many studies have tried to establish the relationship between order degree and GFA from the perspective of the competition between crystallization and vitrification, and have achieved success to a certain extent [33,34]. These studies were based on systems of different components of the same elements, which have similar cluster types. The slight order degree difference between these systems leads to different GFA. For systems with different elements, it is much more complex due to more different cluster types. Disorder degree, such as Voronoi cluster types number, Voronoi cluster variance or Voronoi entropy, cannot be associated with GFA. For example, Al has poor GFA, however, it has a larger disorder degree. While Ta has better GFA, however, it has a largest order degree.

Next, we compared the bond-angle distribution in the simulated monatomic liquids (see Supplementary Figure S2). The overall profile of the bond-angle distribution shows similar feature and temperature-dependent behavior in all liquids, consistent with previous studies [16]. To obtain a quantitative comparison, we analyzed the position of two main peaks in the bond-angle distribution as a function of temperature, which corresponds to two characteristic angles of central atoms with their nearest neighbors. As shown in Figure 5, both peak positions shift to the larger angle with decreasing temperature in these metallic liquids, but less than 60° and 108°, respectively. It can be seen that while both peak angles in liquid Zr are smaller, those in liquid Cu are relatively larger. For liquid Ta, two angles are just smaller than those in liquid Cu. These results are consistent with those obtained in previous studies [35,36]. For HCP, FCC, and BCC structures, the characteristic angles between central atoms and their nearest neighbors are 60°, 90° and 120°, while a perfect icosahedral cluster shows peaks at 63.4°, 116.4° and 180.0° [35]. Thus, it can be seen from Figure 5(a) that the first peak shifts to the right side, gradually approaching 60°. This indicates the formation of close-packed configurations and regular tetrahedra among the nearest-neighbor atoms as temperature decreases. Fig. 5(a) indicates that such configurations could be more populated in liquid Cu, but less in liquid Zr. For the angle associated with the second peak in Fig. 5(b), it is between 102° and 107° in the temperature range, which is much smaller than the characteristic angle of 120° in crystal structures or 116.4° in perfect icosahedral cluster. However, it is very close to the inner angle in a regular pentagonal loop, 108°. This implies that atoms together with their some nearest neighbors tend to form the fragments of pentagonal loop in these monatomic liquids with decreasing temperature. Such tendency is stronger in liquid Cu than others, as shown in Fig. 5(b).

**C. Honeycutt-Anderson Index**

To further quantify the configurations associated with pentagon, we performed Honeycutt-Anderson (HA) Index analysis [16,37], which characterizes the configurations of a pair of two nearest-neighbor atoms with their common neighbor atoms by four integers *ijkl*. Here *i* in HA index denotes the pair of two nearest neighbor atoms; *j* represents the number of nearest neighbors shared in common by the two atoms; and *k* is the number of bonds formed among the shared neighbors. *l* is used to distinguish topologies with the same *ijk* but different packings. For simplicity, only first three indexes were considered in our analysis. We typically took the first minimum of RDFs as the distance cutoff to determine whether a bond is formed between two nearest common atoms [38]. Figure 6 shows the evolution of various HA indices with decreasing temperature in the simulated monatomic liquids. It can be seen that each HA index shows similar temperature dependent behavior in all simulated liquids. While the fraction of 155, 166, and 144 increases, the fraction of other HA indexes decreases with decreasing temperature. However, the fraction of a given HA index varies in these liquids. 155 is the most populated index in all liquids, and it is the highest in liquid Ta. This indicates that pentagon configurations, which is fragment of icosahedral cluster tend to form in quenching liquids. This is consistent with Voronoi cluster analysis. The population of such configurations may be helpful for glass formation.

**D. Analysis of bond-orientational order parameter**

To obtain more comprehensive local atomic structure information, we also analyzed the bond-orientational order (BOO) parameter [39] (see Supplementary Materials S3). Figure S3 shows the distribution of bond-orientational order parameter $Q_4$ and $Q_6$ at various temperatures in the simulated monatomic liquids. While $Q_4$ distribution shifts to the left side, getting narrower and higher as temperature decreases, $Q_6$ distribution exhibit opposite behavior in all simulated liquids. All simulated liquids show similar distributions and temperature evolution, without qualitative difference. Figure 7 shows the change of averaged $Q_4$ and $Q_6$ with decreasing temperature in five monatomic liquids, respectively. It can be seen that in all simulated liquids, averaged $Q_4$ decreases with the decrease of temperature, whereas averaged $Q_6$ increases as temperature decreases. This indicates that the local atomic structures become more ordered with decreasing temperature in metallic liquids. However, the orientational ordering is different in these simulated monatomic liquids. It can be seen that the averaged values of $Q_4$ and $Q_6$ in liquid Cu, Ta and Fe are quite close, indicating that the BOOs in liquid Cu, Ta, and Fe are quite similar. For liquid Zr, it shows smaller values of $Q_6$ and larger values of $Q_4$, compared to other liquids. This indicates that liquid Zr possesses relatively lower bond-orientational ordering.

**E. Characterization of tetrahedral configurations**

To characterize the dense atomic packing feature in these monatomic liquids, we analyzed the tetrahedral configurations in these liquid structures [36,40,41]. In liquid structures, tetrahedral structures formed by four nearest neighbor atoms may not be regular, but distorted. Here a tetrahedral distortion parameter was adopted to characterize the degree of deviation from the regular tetrahedron, which is defined as $\delta = e_{max}/e_{avg} - 1$, where $e_{max}$ and $e_{avg}$ are the longest and average edge lengths in a tetrahedron, respectively [36]. Figure 8(a-e) show the distribution of the tetrahedral distortion parameters and its temperature evolution in these monatomic liquids. It can be seen that the distributions in liquid Cu, Ta, and Fe are similar, but different from those in liquid Al and Zr that are almost the same. The distributions in all simulated liquids show a peak at high temperatures, and the peak position is located around $\delta \sim 0.24$. As temperature decreases, the peak intensity decreases, but peak position keeps unchanged. For liquid Cu, Ta, and Fe, the peak intensity decreases significantly, in contrast to liquid Al and Zr. Meanwhile, a new peak gradually emerges around $\delta \sim 0.1$ with decreasing temperature in liquid Cu, Ta, and Fe. The peak intensity in liquid Ta is higher. In liquid Al and Zr, however, only a shoulder shows up around $\delta \sim 0.08$ in tetrahedral distortion distribution. The distribution features indicate that tetrahedral configurations in all liquids evolve towards regular tetrahedron with decreasing temperature. While tetrahedral structures are more regular in liquid Ta, those in liquid Al and Zr are more distorted. This suggests that liquid Ta is more densely packed in atomic structures, which can be seen more clearly in Fig. 8(f) which shows change of the averaged $\delta_{ave}$ with temperature in these monatomic liquids.

With decreasing temperature, while $\delta_{ave}$ of liquid Ta reduces significantly, the change of $\delta_{ave}$ is relatively small in liquid Zr. Moreover, $\delta_{ave}$ of liquid Ta is much smaller than that in liquid Zr, indicating that tetrahedral configurations in liquid Ta are more regular. For liquid Cu and Fe, $\delta_{ave}$ values are almost the same, and the change trend is quite close to that of liquid Ta. However, $\delta_{ave}$ values are larger than those of liquid Ta. Thus, the above results clearly show the densely packed degree of atomic structures in these monatomic liquids, which is quite useful for understanding GFA of monatomic liquids.

**F. Connection of Atomic packing clusters by sharing common neighbors**

Next, we analyzed the connection of atomic clusters by sharing common neighbor atoms to characterize the atomic packing feature beyond the short-range length scales. The connection of atomic clusters can be evaluated by calculating the number of shared neighbor atoms in two atomic clusters, as schematically illustrated in Figure 9(f), denoted as 1-atom, 2-atom, 3-atom, and 4-atom connection, respectively [28,382-47]. Fig. 9(a-e) show the change in the fraction of cluster connections with decreasing temperature in the simulated monatomic liquids. the connections exhibit the same temperature dependence in liquid Cu, Ta, and Fe, which is different from those in liquid Al and Zr where the connections show the same behavior. This indicates that while liquid Cu, Ta, and Fe possess similar atomic packing feature beyond short-range length scale, liquid Al and Zr contain almost the same connections in neighboring clusters. As shown in Fig. 9, the fraction of 1-atom connection increase with the decrease of temperature in all simulated liquids, but the fraction of 4-atom connection decreases. For 3-atom connection, the fraction increases in liquid Cu, Ta, and Fe, but is almost independent of temperature in liquid Al and Zr. For liquids in which second-peak split occurs, the positions of the two peaks after splitting consist mainly of cluster packings of 1- and 3-atom connection [42]. So, the greater the difference between the fraction of 2-atom connection with 1 and 3, the more obvious the splitting of the second peak. For Cu with the second peak splitting, we found that the fraction of 1- and 3-atom connections increases significantly, and the fraction of 1-atom connection is much higher than that of 2-atom. The splitting of the second peak is not observed for the other four elements. The connection in liquid Ta is very similar to that of Cu, but the 1-atom connection fraction is not high enough to cause splitting. The fraction of 3-atom connection does not change and 1-atom connection are small in Al and Zr. In Fe, the fractions of clusters have some changed except 3-atom cluster connections. Although the second peak splitting of the RDF occurs in liquid Cu, which exhibits certain characteristics of MG, the GFA of Cu is not the best of the five liquids. This suggests that the increase in MRO is only a structural feature and the GFA is not correlated [48].

**E. Dynamical property analysis**

We compared dynamical properties and temperature evolution in these monatomic metallic liquids. Figure 10 (a,b) show diffusion constant $D$ and $\alpha$-relaxation time $\tau_\alpha$ as a function of temperature in these metallic liquids, respectively. Here $D$ and $\tau_\alpha$ were determined by mean-square displacement (MSD) and self-intermediate scattering function, respectively [49-53] (see Supplementary

Materials S4). It can be seen in Fig. 10(a) that $D$ decreases with the decrease of temperature in these monatomic liquids. $D$ is quite similar in liquid Cu and Ta, and smaller than that in other liquids. On the other hand, liquid Zr and Al show almost the same $D$ values. For structural relaxation in these monatomic liquids, $\alpha$-relaxation time $\tau_\alpha$ is larger in liquid Ta, and increases more significantly with decreasing temperature. It can be seen that $\tau_\alpha$ in liquid Cu is slightly smaller than that in liquid Ta. In liquid Al and Zr, the increase of $\tau_\alpha$ is also relatively slow. The above results indicate that dynamics is much slower in liquid Ta, but faster in liquid Al and Zr. The slower dynamics in liquid Ta benefits the glass formation. As shown above, liquid Ta contains more ordered atomic structures with more icosahedral-like configurations and more densely atomic packings, compared to other monatomic liquids. All these atomic structure characteristics may facilitate slower dynamics, favoring GFA.

Fig. 10(c) shows the parametric relation of $D$ as a function of $\tau_\alpha/T$ in monatomic liquids. It is shown that $D \sim \tau_\alpha/T$ obeys Stokes-Einstein (SE) relation, that is, $D \propto (\tau_\alpha/T)^{-1}$ in high temperature liquids. With the decrease of temperature below a certain point, the simulated monatomic liquids gradually deviate from the linear lines with slope of -1, respectively, indicating the breakdown of SE relation and governing of fractional SE relation in liquid dynamics at lower temperatures, that is, $D \propto (\tau_\alpha/T)^{-\kappa}$ (0<$\kappa$<1) [54,55]. The behavior of $D \sim \tau_\alpha/T$ does not show particular features in these monatomic liquids. However, the crossover temperature from SE to fractional SE relation varies in these liquids, which is about $0.95T_m$, $0.97T_m$, $0.93T_m$, $0.93T_m$, and $0.89T_m$ in liquid Cu, Ta, Fe, Al, and Zr, respectively. This indicates that the decoupling of diffusion constant and relaxation time in liquid Ta occurs just below melting temperature. In liquid Zr, however, the crossover temperature is as low as $0.89T_m$. Thus, the crossover temperature could be related to GFA of liquids.

We also analyzed the evolution of dynamic heterogeneity with decreasing temperature in these monatomic liquids. Here non-Gaussian parameter $\alpha_2(t)$ was calculated to quantify the dynamical heterogeneity in liquids defined as [56,57]

$$\alpha_2(t) = \frac{3\langle \Delta r^4(t) \rangle}{5\langle \Delta r^4(t) \rangle^2} - 1$$

where $\Delta r$ is the atomic displacement in the time interval of $t$. Figure S5 shows the typical behavior of $\alpha_2(t)$ at different temperatures in these monatomic liquids, which exhibits a pronounced peak at a characteristic time scale of $t_{max}$. This means that the degree of dynamical heterogeneity reach maximum $\alpha_2^{max}$ at $t_{max}$. To characterize the growing dynamical heterogeneity, the values of $\alpha_2^{max}$ was examined a s function of temperature. Figure 11 shows the temperature dependence of $\alpha_2^{max}$ and $t_{max}$ in these monatomic liquids. The values of $\alpha_2^{max}$ and $t_{max}$ are quite small in high-temperature liquids. With decreasing temperature, $\alpha_2^{max}$ and $t_{max}$ increases quickly, especially in liquid Ta. This indicates that liquid dynamics in Ta becomes more heterogenous, compared to other monatomic liquids. The values of $\alpha_2^{max}$ is almost the same in liquid Cu and Ta. However, $t_{max}$ in liquid Ta is much larger than that in liquid Cu, indicating that dynamical heterogeneity in liquid Ta is more significant. As shown in Fig. 11, liquid Zr exhibit much less heterogenous dynamics. This indicates that dynamical heterogeneity in liquids may also play a certain role in GFA.

**IV. SUMMARY**

We systematically studied the structures and dynamics of five common metallic liquids. It is found that the liquids arrange more regularly with the decrease of temperature. A more ordered atomic structure and more icosahedral-like structures may favor GFA. In dynamics, SE relationship for liquids breaks down gradually with the temperature decreasing, and the crossover temperature could be related to GFA of liquids. Dynamical heterogeneity in liquids may also play a certain role in GFA. Through the systematic study of various structural quantities of these systems, it deepens understanding of monatomic liquid.

**Acknowledgements**

This work was supported by National Natural Science Foundation of China (Nos. 52031016 and 51631003).

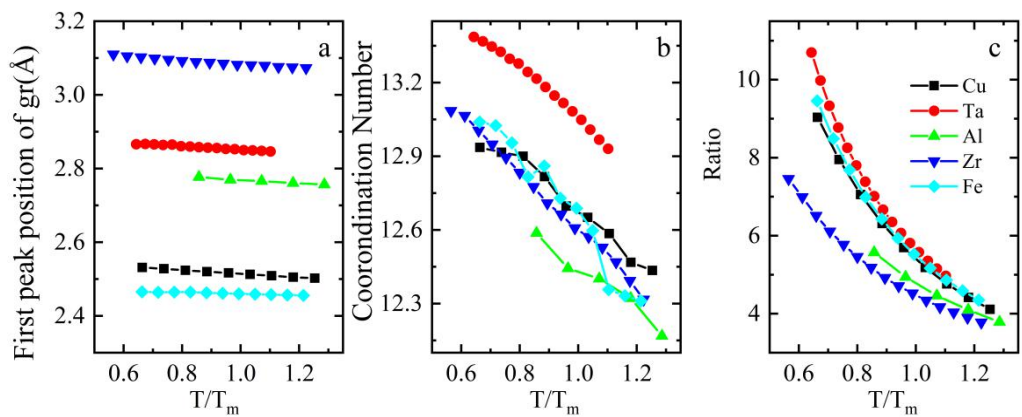

Figure 1 The first peak position of radial distribution functions (RDFs) $g(r)$ (a), coordination number (b), and ratio of the first peak height to the first local minimum in RDFs (c) as a function of temperature in simulated monatomic liquids, respectively.

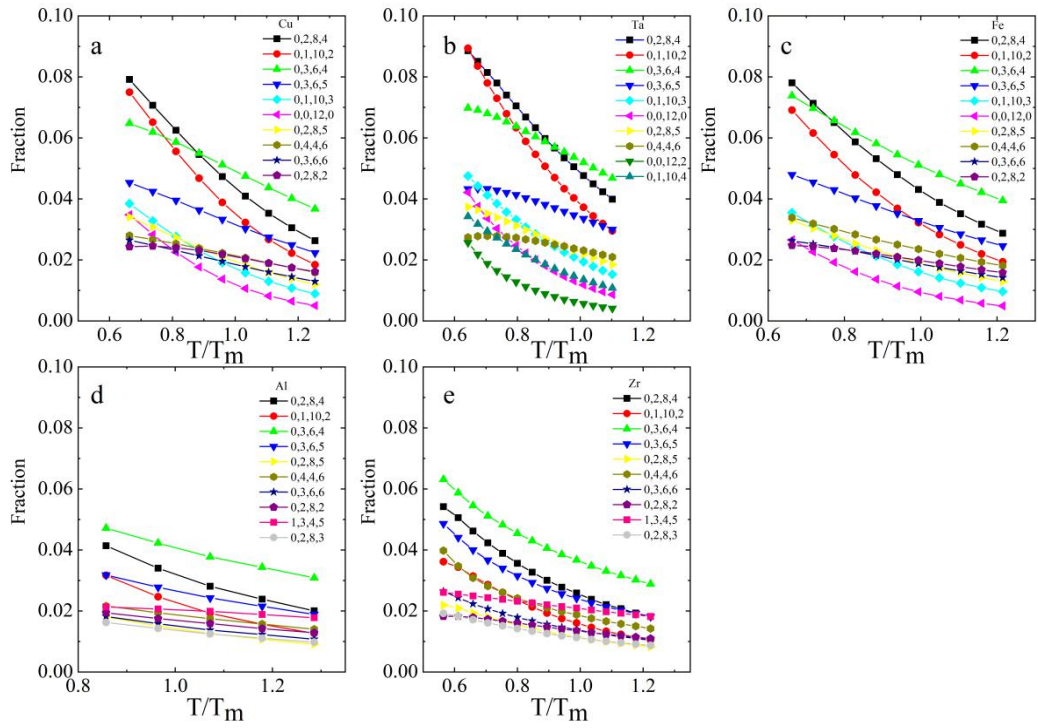

Figure 2 Evolution of the fraction of top 10 populated Voronoi clusters with decreasing temperature in five simulated monatomic liquids, respectively.

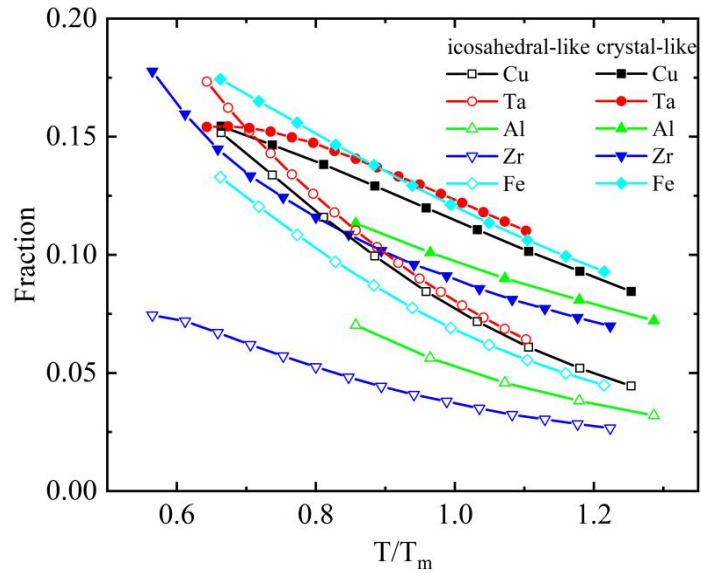

Figure 3 Evolution of the fraction of crystal-like (<0,3,6,4>, <0,3,6,5>, <0,4,4,6>, <0,4,4,7>, <0,6,0,8>, <0,5,2,8>) (a) and icosahedral-like (<0,0,12,0>, <0,1,10,2>, <1,0,9,3>, <0,2,8,2>) (b) clusters with decreasing temperature in five simulated monatomic liquids, respectively.

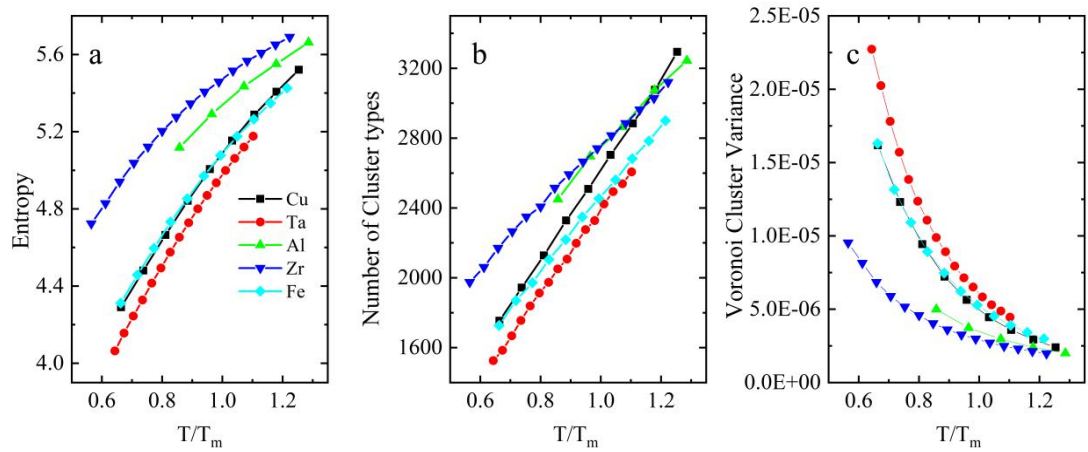

Figure 4 Evolution of Voronoi entropy (a), number of cluster types (b), and variance of Voronoi cluster fraction $\sigma^2$ (c) with decreasing temperature in five monatomic liquids, respectively.

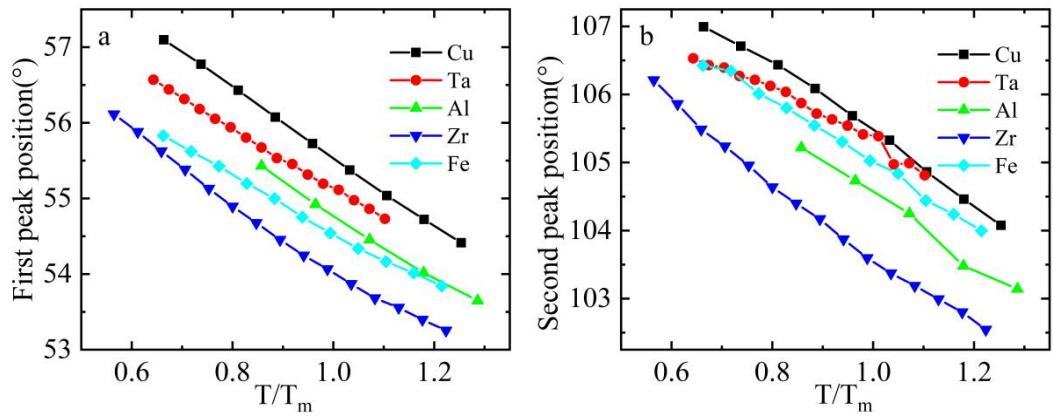

Figure 5 Variation of the first (a) and second (b) peak position in bond-angle distribution with temperature in five monatomic liquids, respectively.

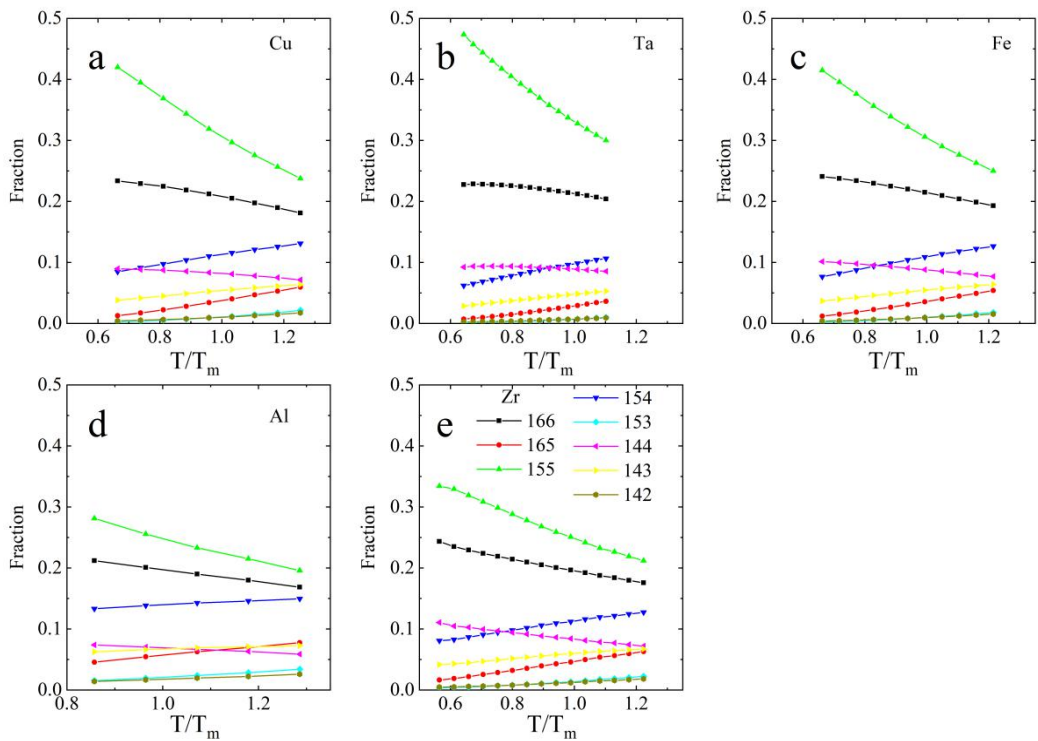

Figure 6 Evolution of HA indexes with decreasing temperature in five monatomic liquids, respectively.

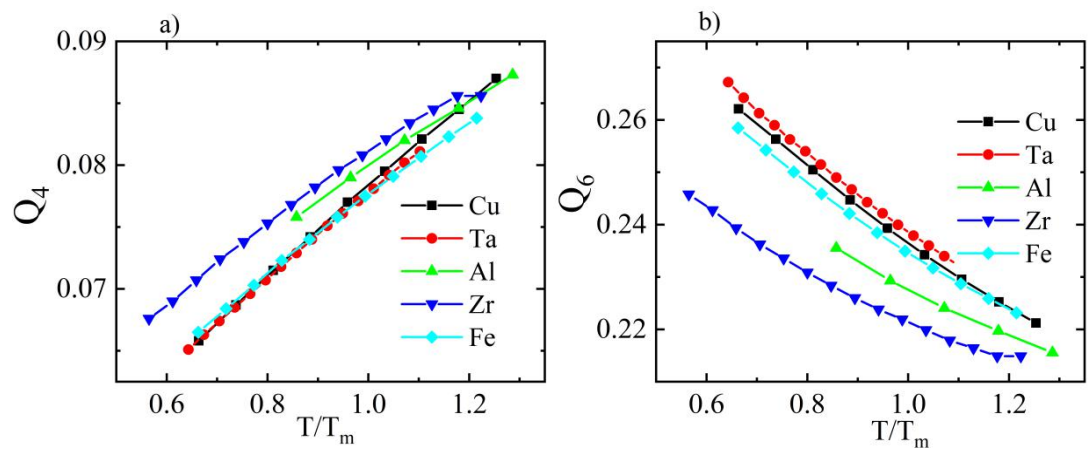

Figure 7 Evolution of bond-orientational order parameter $Q_4$ (a) and $Q_6$ (b) with decreasing temperature in five monatomic liquids, respectively.

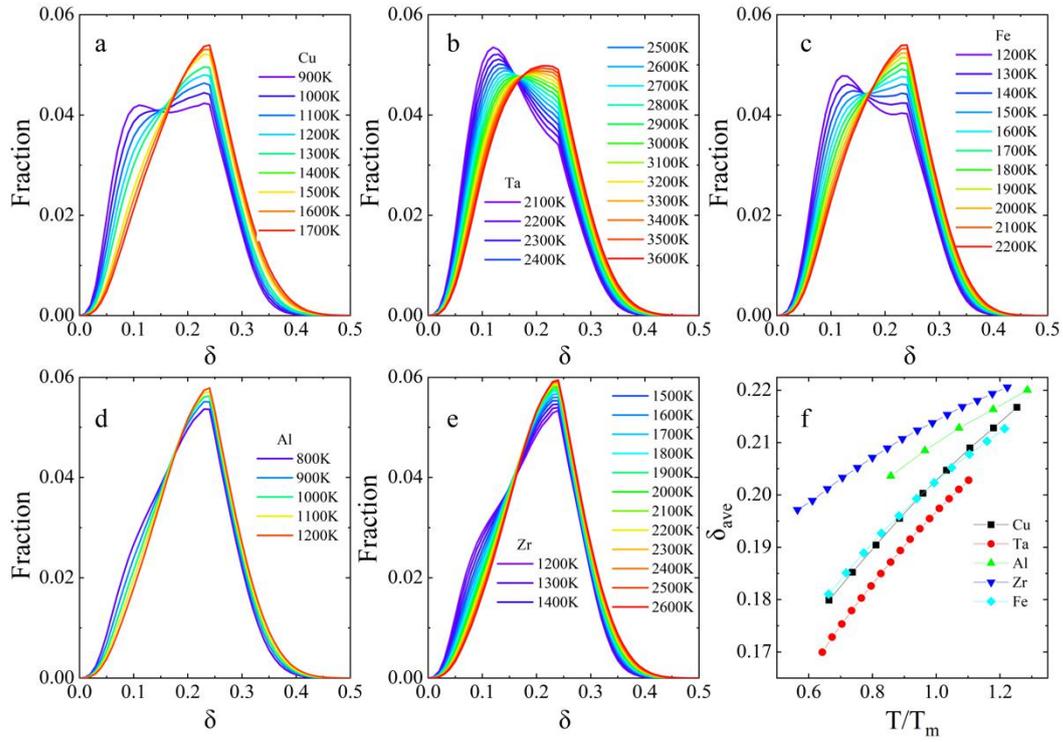

Figure 8 Distribution of tetrahedral distortion parameter in five monatomic liquids at various temperatures (a-e), respectively. (f) Averaged tetrahedral distortion parameter $\delta_{ave}$ as a function of temperature in these monatomic liquids.

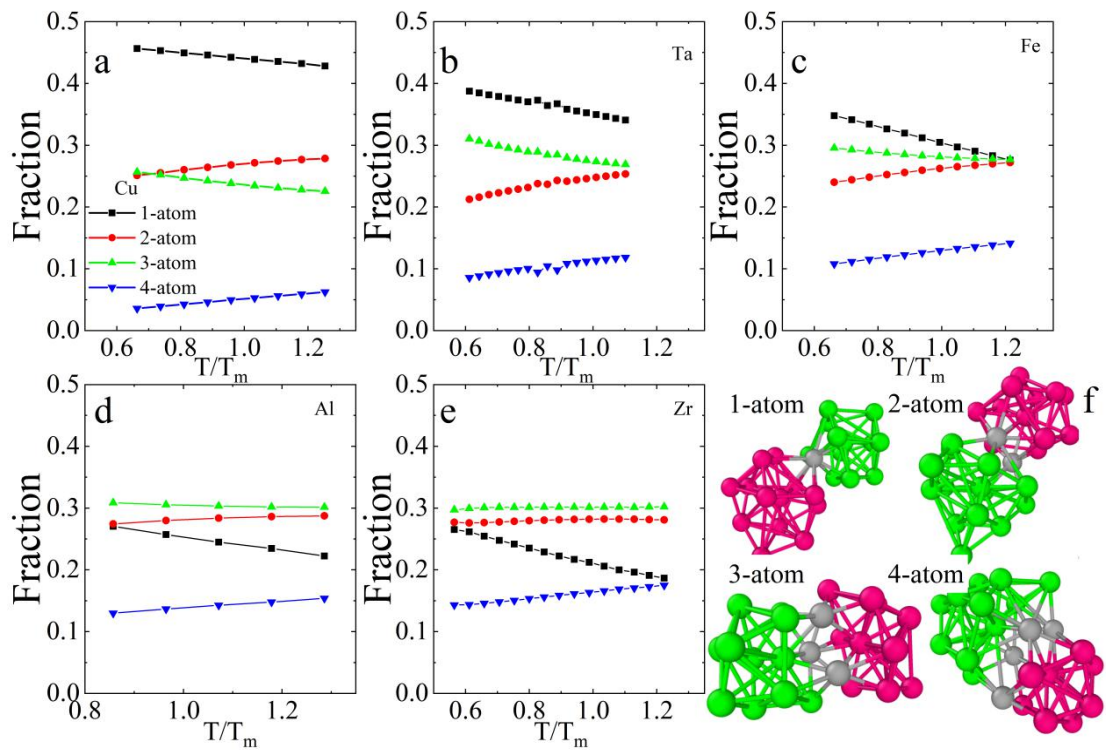

Figure 9 (a-e) Evolution of fraction of 1-atom, 2-atom, 3-atom and 4-atom connection by with decreasing temperature in five monatomic liquids, respectively; (f) Schematic of cluster connection in the second-nearest neighbor atoms.

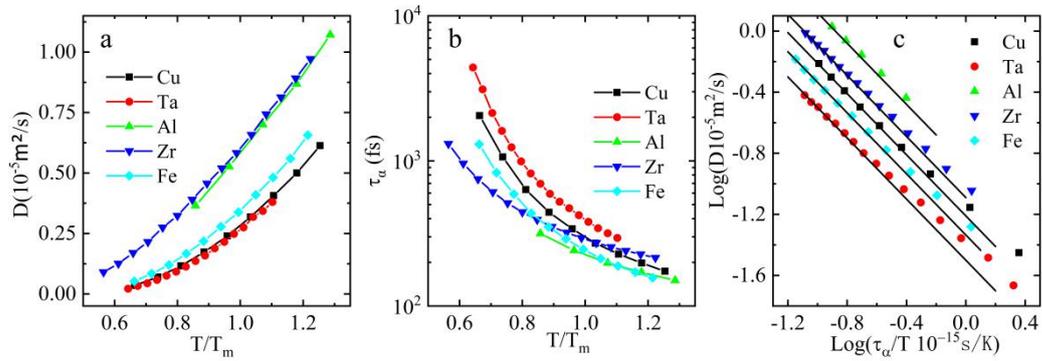

Figure 10 Temperature dependence of diffusion constant $D$ (a) and $\alpha$-relaxation time $\tau_\alpha$ (b) in the simulated monatomic liquids, respectively. (c) The parametric relation of $D$ as a function of $\tau_\alpha/T$ in five monatomic liquids. Solid symbols denote simulation data, and solid lines are linear curves with slope of -1.

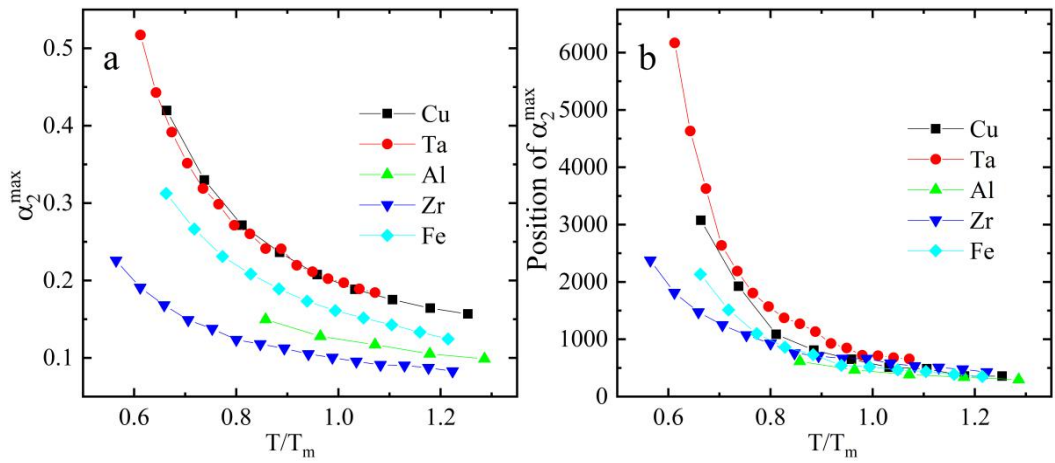

Figure 11. Temperature dependence of the maximum value of non-Gaussian parameter $\alpha_2^{max}$ (a) and the time scale of $\alpha_2^{max}$ in the simulated monatomic liquids, respectively.

# Supplementary material for

## Atomic structural characteristics and dynamical properties in monatomic metallic liquids via molecular dynamics simulations

X. Qin, J.Q. Wu, and M. Z. Li

*Department of Physics and Beijing Key Laboratory of Opto-Electronic Functional Materials & Micro-Nano Devices, Renmin University of China, Beijing 100872, China*

**S1. Radial distribution function**

In our studies, radial distribution function (RDF) in the simulated monatomic metallic liquids were calculated according to

$$g(r) = \frac{1}{4\pi r^2 \rho N} \sum_{i=1}^{N} \sum_{j=1, j \neq i}^{N} \delta(\vec{r} - |\vec{r}_{ij}|)$$

which characterizes the probability of finding atoms at distance $r$ from an atom [1]. Here $N$ is the total number of atoms, $\rho$ the average number density of a liquid, and $|\vec{r}_{ij}|$ the distance between atom $i$ and $j$, respectively. Figure S1 shows RDFs of five simulated liquids at various temperatures. Typical behavior and temperature evolution of RDFs in liquids can be observed.

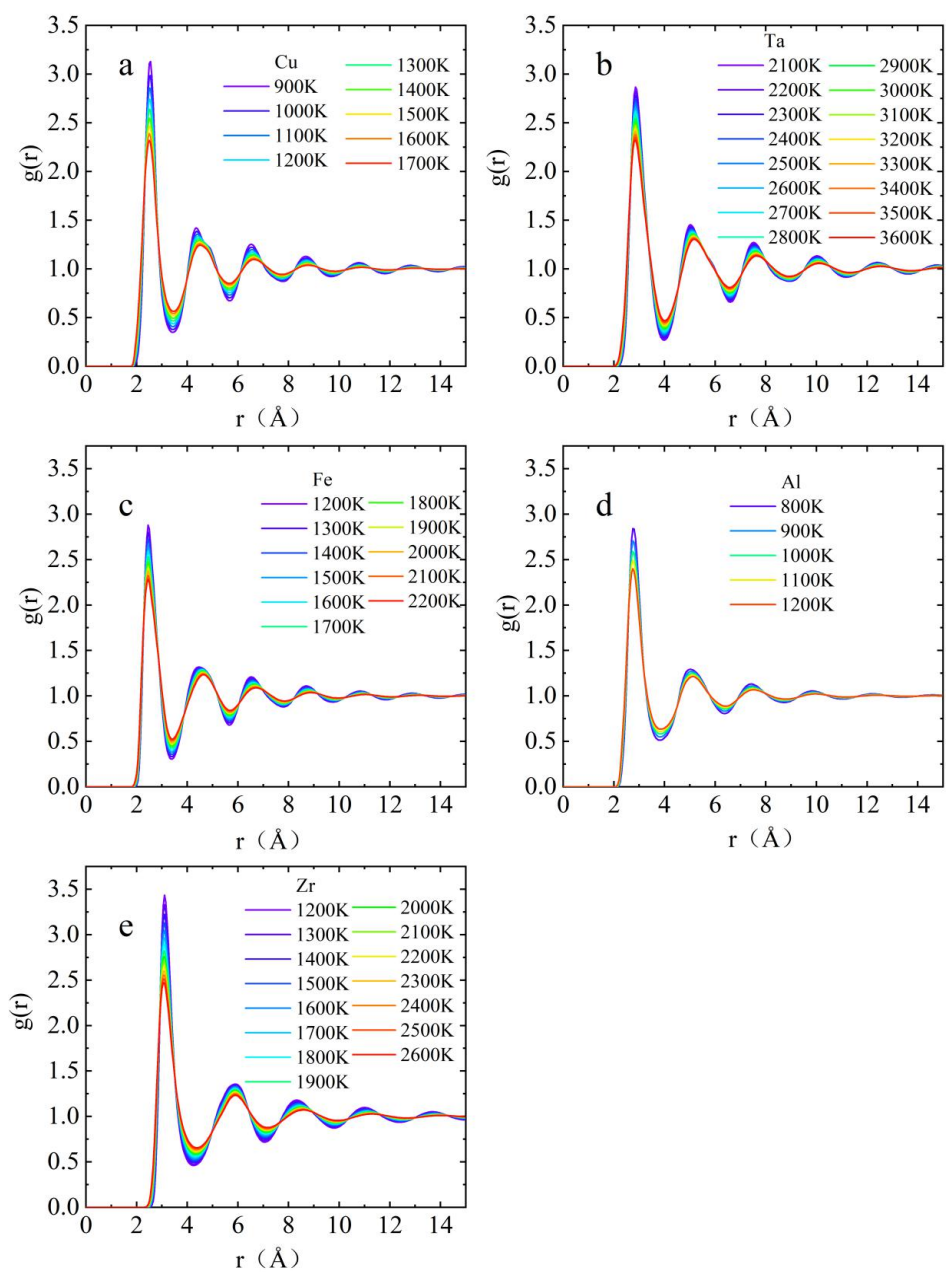

Figure S1. Radial distribution functions g(r) at various temperatures in the simulated monatomic liquids.

## S2 Bond-angle distribution

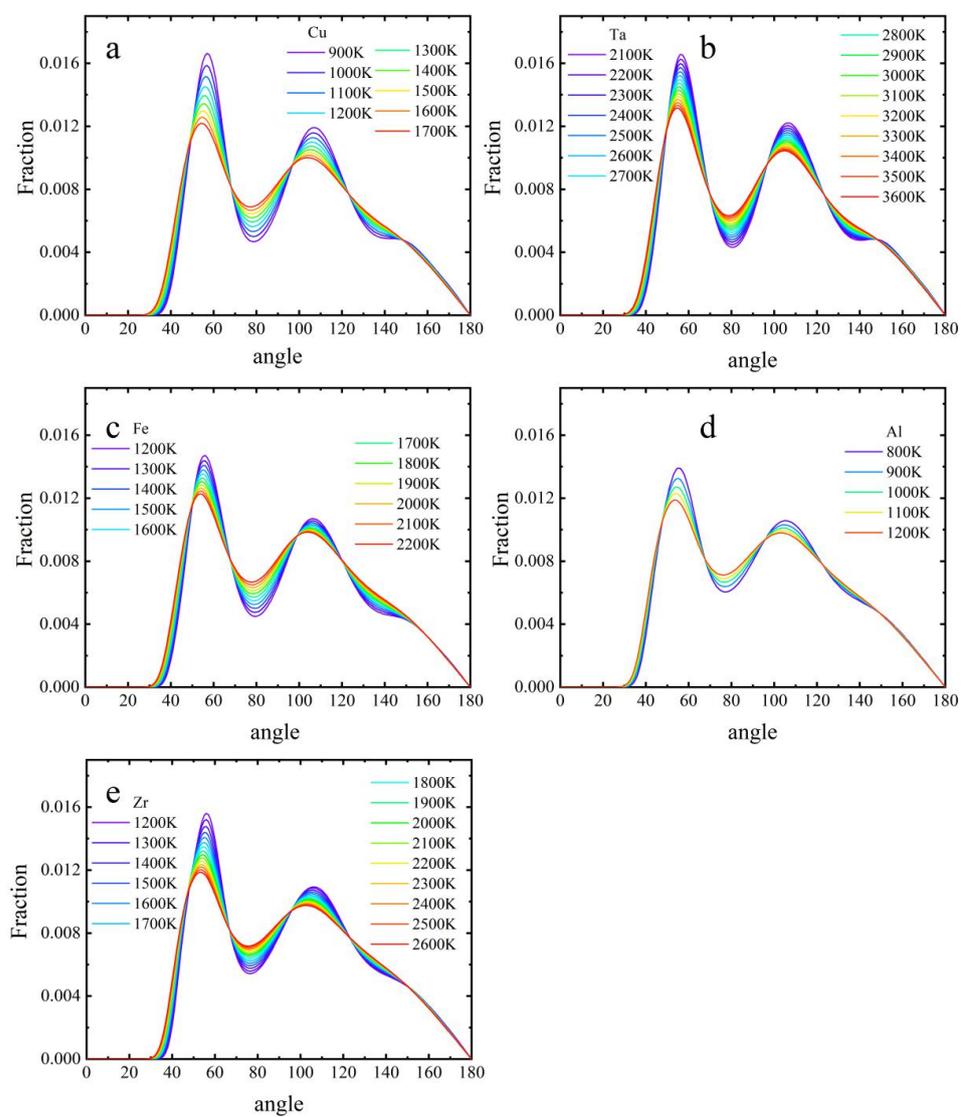

Figure S2. Bond-angle distributions at various temperatures in the simulated monatomic liquids.

S3. Bond-Orientational order (BOO) parameter

The bond-orientational order (BOO) parameter was developed by PJ Steinhardt et al. in 1983 [2], and is often used to characterize the atomic ordering in local structures in disordered materials. The average BOO parameter for an Voronoi cluster can be defined as

$$\overline{Q}_{lm}(\vec{r}_i) = \frac{1}{N_b(i)} \sum_{j=1}^{N_b(i)} Y_{lm}(\theta(\vec{r}_i), \phi(\vec{r}_i))$$

where $Y_{lm}$ is the spherical harmonic function of degree $l$, and $m$ is an integer running from $m=-l$ to $m=l$. $\theta(\vec{r}_i)$ and $\phi(\vec{r}_i)$ are the polar angles of the bond measured with respect to some reference coordinate system. $N_b(i)$ denotes the number of nearest neighboring atoms around atom $i$.

The rotationally invariant combinations can be expressed as

$$Q_l(\vec{r}_i) = \sqrt{\frac{4\pi}{2l+1} \sum_{m=-l}^{l} \overline{Q}_{lm}^*(\vec{r}_i) \overline{Q}_{lm}(\vec{r}_i)}$$

Figure S3 shows the distribution of bond-orientational order parameter $Q_4$ and $Q_6$ at various temperatures in the simulated monatomic liquids. $Q_4$ decreases with the decrease of temperature, and $Q_6$ increases with the decrease of temperature.

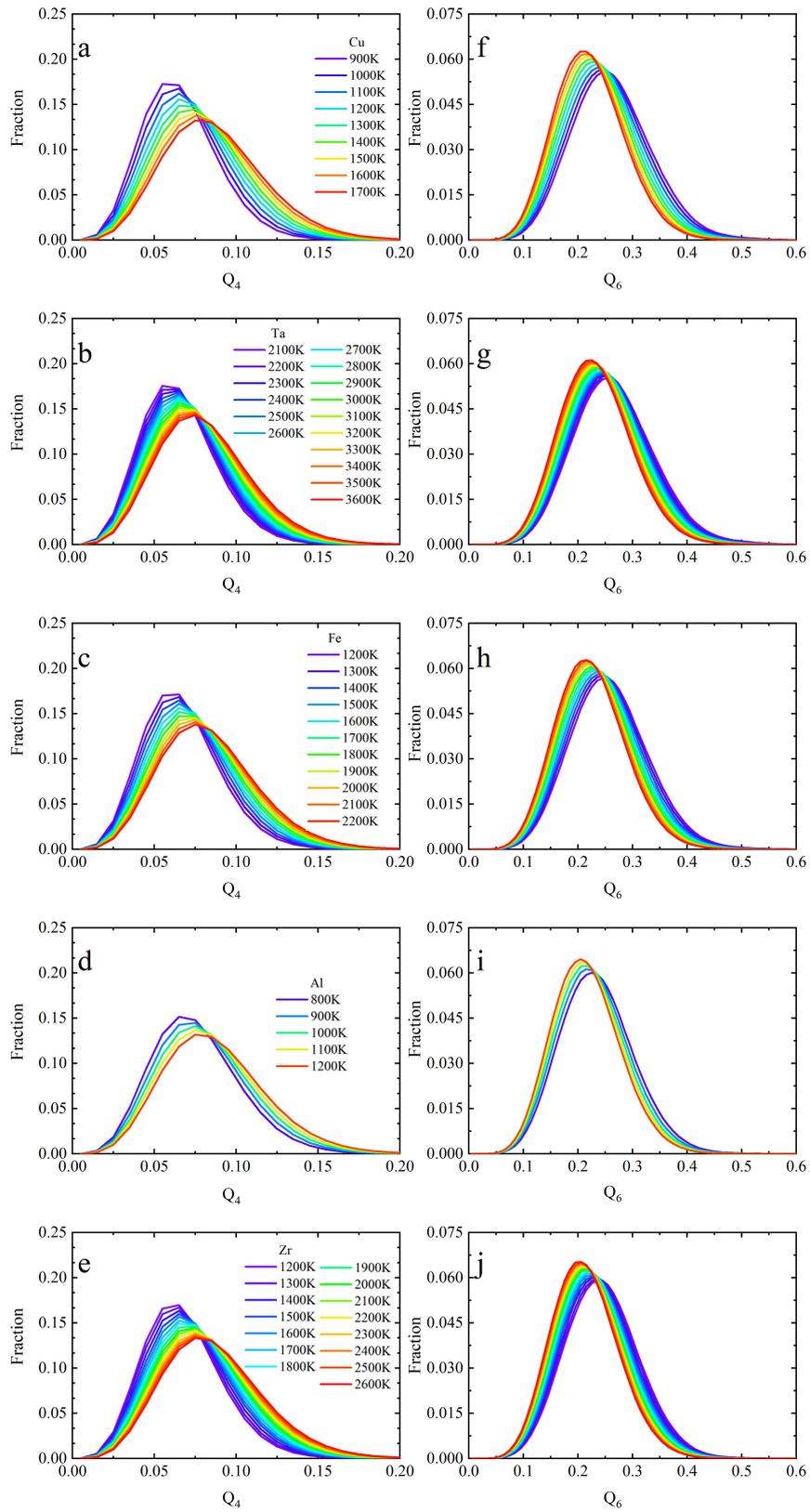

Figure S3. Distribution of bond-orientational order parameter $Q_4$ (a-e) and $Q_6$ (f-j) at various temperatures in the simulated monatomic liquids.

## S4. Mean-square displacement and self-intermediate scattering function

In our work, mean-square displacement (MSD) and self-intermediate scattering function (SISF) were calculated by

$$\langle \Delta r_i^2(t) \rangle = \frac{1}{N} \sum_{i=1}^{N} \langle |\vec{r}_i(t) - \vec{r}_i(0)|^2 \rangle$$

and

$$F_S(q, t) = \frac{1}{N} \langle \sum_{i=1}^{N} \exp[i\vec{q} \cdot (\vec{r}_i(t) - \vec{r}_i(0))] \rangle$$

respectively, where $N$ is the number of atoms, $\vec{r}_i(t)$ the atomic position of atom $i$ at time $t$, and $\vec{q}$ the wave vector often fixed at $q_{max} = |\boldsymbol{q}|$ of the first peak position in structure factor, respectively [3,4]. Thus, self-diffusion coefficient of the system can be determined according to MSD by [5],

$$D = \lim_{t \to \infty} \frac{\langle \Delta r_i^2(t) \rangle}{6t}$$

For $\alpha$-relaxation time $\tau_\alpha$, it is often evaluated by the time scale when SISF decays to $e^{-1}$ of its initial value [6]. Figure S4 shows temperature evolution of MSD and SISF in simulated monatomic metallic liquids, respectively.

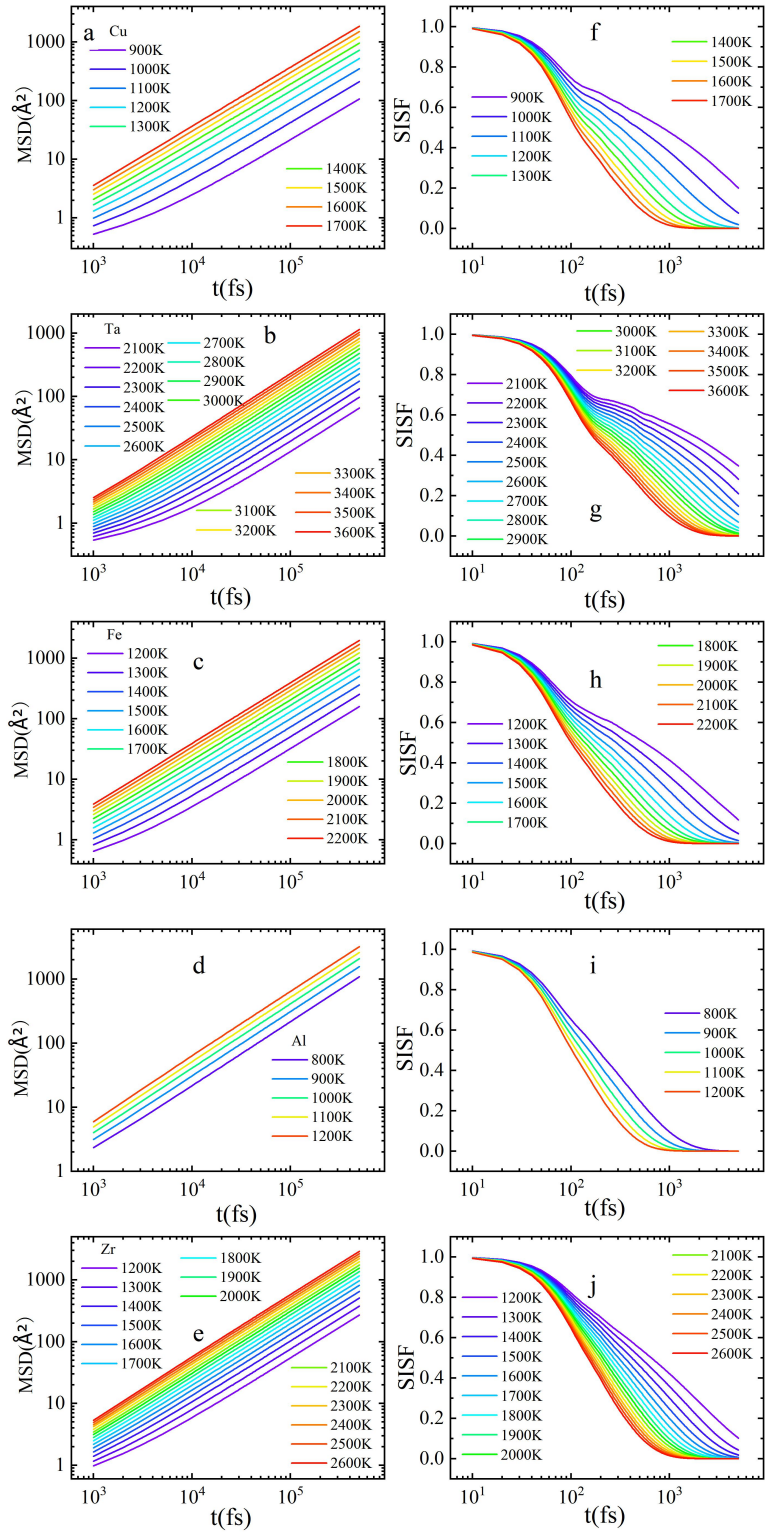

Figure S4. Temperature evolution of mean-square displacement (MSD) and self-intermediate scattering function (SISF) in the simulated monatomic metallic liquids, respectively..

## S5. Non-Gaussian parameter

Non-Gaussian parameter is often used to characterize dynamic heterogeneity in glass-forming liquid in cooling process. It is defined as

$$\alpha_2(t) = \frac{3\langle \Delta r^4(t) \rangle}{5\langle \Delta r^4(t) \rangle^2} - 1$$

where $\Delta r(t)$ is the atomic displacement in the time interval of $t$. $\langle \rangle$ denotes the ensemble average.

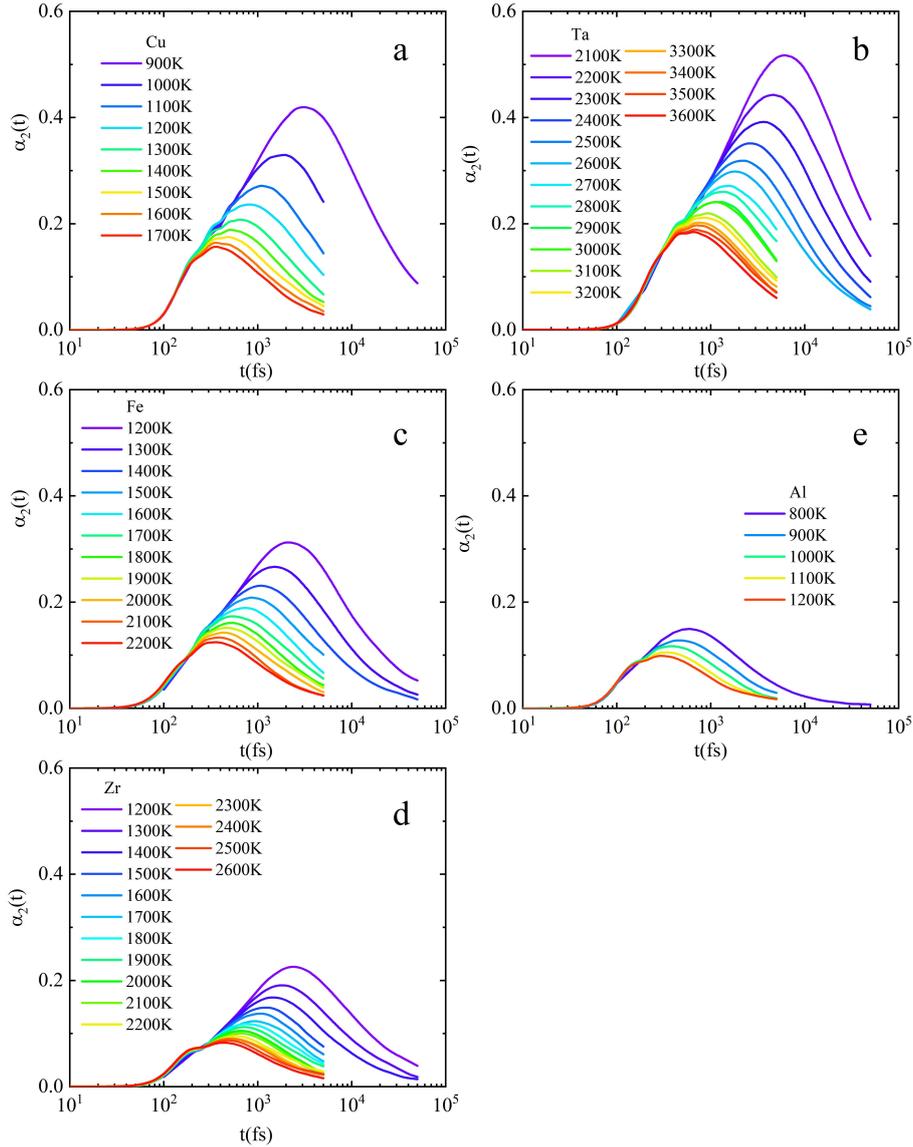

Figure S5. Evolution of non-Gaussian parameter with temperature in the simulated monatomic liquids, respectively.